\documentclass[10pt,letterpaper]{article}
\usepackage{opex3}
\usepackage{graphicx}
\usepackage{color}
\usepackage{cite}

\begin{document}
\begin{sloppy}

\title{Discrete dissipative localized modes in nonlinear magnetic metamaterials}

\author{Nikolay N. Rosanov$^{1,2}$, Nina V. Vysotina$^{2}$, Anatoly N. Shatsev$^{2}$, \\Ilya V. Shadrivov$^{1,3}$,  David A. Powell$^3$, and Yuri S. Kivshar$^{1,3}$}

\address{
$^1$National Research University of Information Technologies, Mechanics and Optics (ITMO), \\ St. Petersburg 197101, Russia\\
$^2$ S.I. Vavilov State Optical Institute, St. Petersburg 199034, Russia\\
$^3$Nonlinear Physics Centre, Research School of Physics and Engineering, Australian National University, Canberra ACT 0200, Australia}

\begin{abstract}
We analyze the existence, stability, and propagation of dissipative discrete localized modes in one- and two-dimensional nonlinear lattices
composed of weakly coupled split-ring resonators (SRRs) excited by an external electromagnetic field. We employ the near-field interaction approach
 for describing quasi-static electric and magnetic interaction between the resonators, and demonstrate the crucial importance of the electric coupling, which can completely reverse the sign of the overall interaction between the resonators. We derive the effective nonlinear model and analyze the properties of nonlinear localized modes excited in one- and two-dimensional lattices. In particular,  we study nonlinear magnetic domain walls (the so-called switching waves) separating two different states of nonlinear magnetization, and reveal the bistable dependence of the domain wall velocity on the external field. Then, we study two-dimensional localized modes in nonlinear lattices of SRRs and demonstrate that larger domains may experience modulational instability and splitting.
\end{abstract}



\section{Introduction}

Important building blocks of electromagnetic metamaterials~\cite{book} are the split-ring resonators (SRRs) or other types of subwavelength resonant elements which are arranged in one-, two-, or three-dimensional lattices. In general, the response of a metamaterial is not simply given by a sum of the responses of individual resonators, but it depends also on the near-field interaction between the
resonators within the system~\cite{prb,Gorkunov:2002-263:EPJ}. A standard theoretical approach for analyzing the properties of metamaterials is based on the effective medium approximation when the structure is treated as a homogeneous medium characterized by effective macroscopic parameters. This approximation is justified when the characteristic wavelength of the electromagnetic field is much larger than the period of the microstructured medium. However, for smaller wavelength or in the cases when the internal resonant modes are important, the metamaterials demonstrate their discrete nature and strong nonlocal effects~\cite{prb_pavel}, so they should be described as lattices of resonant elements by employing the techniques usually used in the analysis of photonic crystals and waveguide arrays~\cite{book2}.

\begin{figure}[b]
\centering\includegraphics[width=13cm]{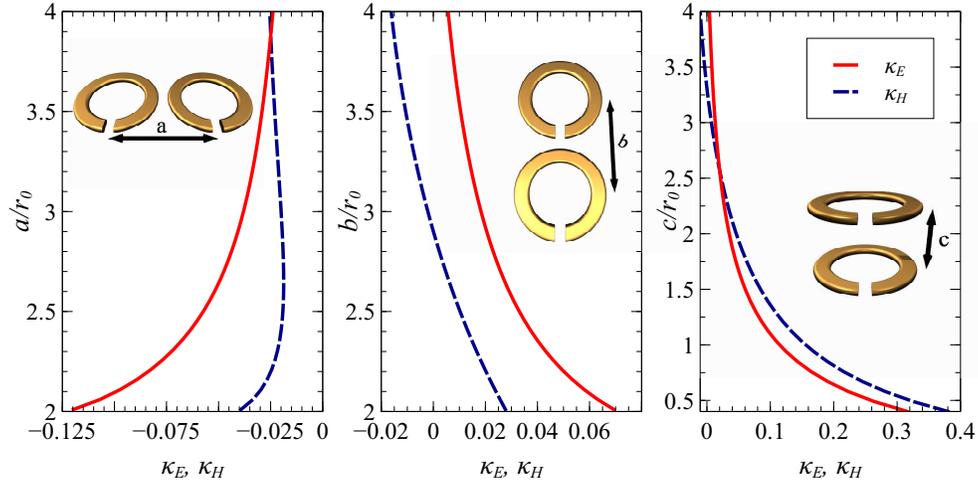}
\caption{Electric and magnetic interaction coefficients (solid and dashed curves, respectively) as functions of the spacing between the ring centers for the normalized frequency $\Omega = -0.2$, when the rings are offset in (a) $x$ direction, (b) $y$ direction, and (c) $z$ -direction. Insets show schematically corresponding ring positions.}
\label{fig1}
\end{figure}

The discrete physics of composite metamaterials can be studied using a novel type of coupled nonlinear equations which describe the so-called nonlinear magnetoinductive waves~\cite{pnfa} when the magnetic response of a metamaterial may become bistable. In this paper, we extend the earlier analysis of Ref.~\cite{pnfa} and consider a lattice of SRRs analyzing various types of nonlinear dissipative discrete localized modes. Although the nonlinear discrete equations were originally derived in Ref.~\cite{pnfa} and further analyzed in Refs.~\cite{tsiron1,tsiron2,tsiron3,tsiron4,cui}, here we introduce an important generalization of those equations which takes into account both electric and magnetic near-field coupling between the neighboring sites, as was recently described by a deeper analysis of resonator interaction~\cite{prb,prb2}.

 We demonstrate that in the nonlinear regime the magnetic response of an array of split-ring resonators may become bistable, so that a nonlinear metamaterial can support the propagation of domain walls (also called {\em switching waves}) separating the regions of different values of the magnetization, and study the motion of such domain walls under the action of an external field, revealing the hysteresis in the dependence of the velocity on the applied field. We also analyze the existence and stability of nonlinear localized modes in two-dimensional magnetic lattices of SRRs including the decay of magnetic domains via modulational instability.

The paper is organized as follows.  Section~2 is devoted to the derivation of the new nonlinear model that takes into account both electric and magnetic interaction between the split-ring resonators. This new model is then applied in Sec.~3 to the study of nonlinear localized modes in dissipative one- and two-dimensional lattices composed of nonlinear SRRs. More specifically, we confirm the existence of magnetic domain walls in arrays of SRRs when both the interaction terms are taken into account (Sec.~3), and also study the nonlinear dynamics of dissipative localized modes in two-dimensional lattices when modulational instability may break up the broader modes into smaller modes (Sec.~4). Finally, Sec.~5 concludes the paper.

\section{Near-field interaction between split-ring resonators}

\begin{figure}
\centering\includegraphics[width=13cm]{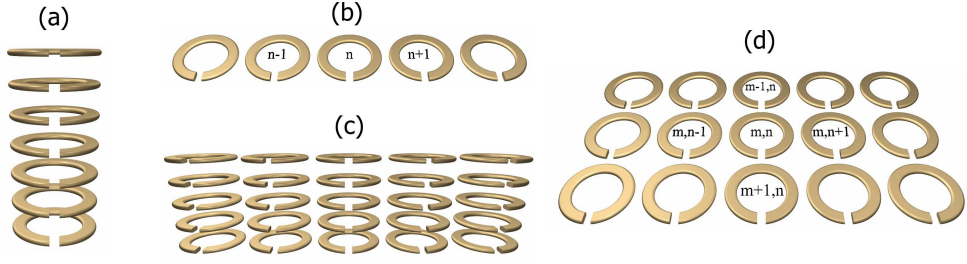}
\caption{Examples of (a,b) arrays and (c,d) two-dimensional lattices of weakly
coupled nonlinear split-ring resonators creating the simplest nonlinear magnetic metamaterials.}
\label{fig2}
\end{figure}

The equation describing the current $I_{n,q,m}$ in a split-ring resonator with indices (n,q,m), corresponding to the axes (x,y,z), has the form~\cite{prb}
\begin{equation}\label{Kirchhoff}
L\frac{dI_{n,q,m}}{dt} + R I_{n,q,m} + \frac{Q_{n,q,m}}{C} = \mathcal{E}_{n,q,m} -  \sum_{\bf j} \left( M_{\bf j} \frac{dI_{\bf j}}{dt}  + P_{\bf j} Q_{\bf j} \right),
\end{equation}
where $L, R, C$ are inductance, resistance and capacitance of the resonator, $Q_{n,q,m}$ is the charge in the resonator, {\bf j} describes indices of the nearest neighboring resonators, $M$ and $P$ are electric and magnetic interaction coefficients, respectively, $\mathcal{E}_{n,q,m}$ is an external electromotive force.

Following~\cite{pnfa}, we assume that the capacitance of the resonator is nonlinear, i.e. it depends on the voltage $U$ across the gap of the resonator: $C_{NL} = C_0 + \Delta C_{NL}\left(|U_{n,q,m})|^2\right)$, where the nonlinear correction to the capacitance is small. Using the slowly varying approximation~\cite{pnfa} we obtain the following equation for the normalized current in the split-ring resonator:
\begin{equation}\label{svea}
i\frac{d\psi_{n,q,m}}{d\tau} - \left( 2 \Omega -i \gamma + \alpha |\psi_{n,q,m}|^2 \right) \psi_{n,q,m}  = \Sigma_{n,q,m} + \sum_{\bf j} K_{\bf j} \psi_{\bf j},
\end{equation}
where the dimensionless variables $\tau = \omega_0
t$, $\omega_0^2 = 1/LC_0$, $\Omega = (\omega - \omega_0)/\omega_0$, and $\Psi_{n,q,m} = {\cal I}_{n,q,m}/{\cal I}_c$, where ${\cal I}_c = \omega_0
C_0 U_c$ is the characteristic nonlinear current, $U_c =
E_c \times d_g$ is the characteristic nonlinear voltage, $\gamma =
R/L\omega_0$ is the damping coefficient, and
\[
\Sigma_{n,q,m} = - \frac{\omega H_0 \pi r_0^2}{c \omega_0 L {\cal I}_c},
\]
 is the normalized
electromotive force, ${\cal
I}_{n,q,m}$ is the amplitude of the harmonic current $I_{n,q,m}$. The effective interaction coefficients
$K_{\bf j} = \kappa_{{\bf j}, H}\omega/\omega_0 -  \kappa_{{\bf j}, E}\omega_0/\omega$,
where normalized electric and magnetic interaction coefficients are $ \kappa_{{\bf j}, H} = M/L$ and $\kappa_{{\bf j}, E} = P/L\omega_0^2$. These interaction coefficients crucially depend on the mutual position of the resonators, and the distributions of currents and charges in conductors~\cite{prb}.

\begin{figure}
\centering\includegraphics[width=11cm]{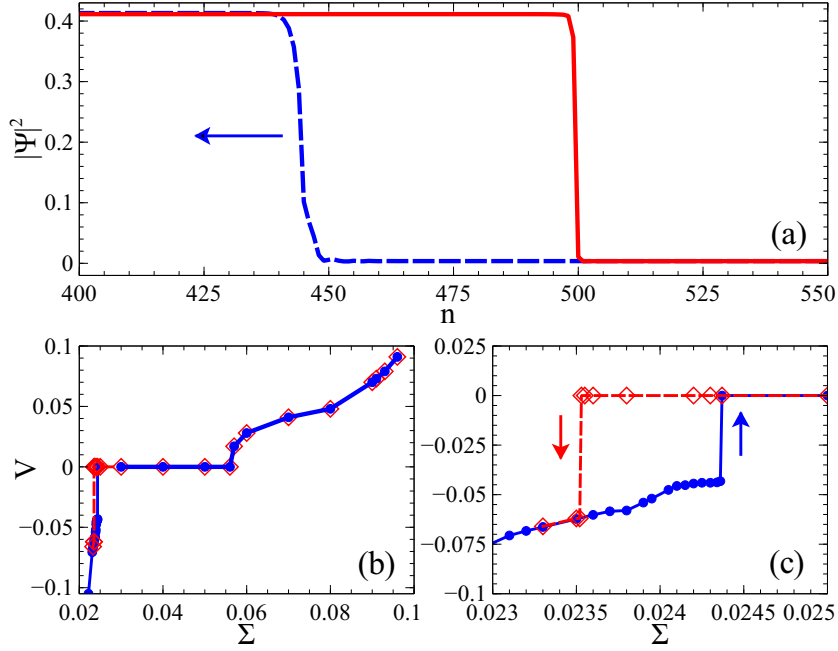}
\caption{(a) Dependence of the magnetization on the index {\em n} for $\Sigma$ = 0.024. Solid - standing, and dashed - moving domain wall, corresponding to the top and bottom branches on the panel (c) in this Figure. (b) Dependence of the velocity of the domain wall as a function of the external excitation $\Sigma$. (c) shows close up of the bistable region in panel (b), with arrows indicating directions of the jumps. All results are obtained for $\Omega = -0.2$, $K_z = -0.02$.}
\label{fig3}
\end{figure}

To be more specific, in further calculations we assume that the resonators have the single ring geometry identical to that used in Ref.~\cite{prb}, namely: rings have average radius $r_0$=2.25 mm, track width of 0.5 mm, metal thickness of 0.03 mm, gap width of 1 mm. For such resonators, the interaction coefficients were calculated using our approach presented in Ref.~\cite{prb}, where we have also taken into account the effect of retardation~\cite{Mark}. The calculated coefficients are presented in Fig.~(\ref{fig1}). We verified that the resonant frequencies of a pair of interacting rings described by the interaction coefficients match the resonances found from direct numerical simulations using commercial software CST Microwave Studio. This makes us confident that the coefficients can {\em quantitatively} describe near-field interaction of the resonators. Importantly, the effective interaction coefficient $K_{\bf j}$ in Eq.~\ref{svea} depends on the difference between electric and magnetic coefficients, which for experimentally realistic parameters may become comparable. In contrast to the previous works, which assumed only approximate magnetic interaction between the resonators, when the interaction of in-plane resonators is identical in x- and y-directions, in our case we see from the Figs.~\ref{fig1} (a,b) that the interaction is strongly different in x- and y-direction. Remarkably, the corresponding effective interaction coefficients $K$ can have different signs.

\section{Discrete dissipative localized modes}

\subsection{Domain walls in SRR arrays}

First, we study switching wave propagation in one-dimensional arrays of split-ring resonators, see Figs.~\ref{fig3}, extending our previous analysis~\cite{our_JETPL}. We select the geometry shown in Fig.~\ref{fig2}(a) with the spacing between the resonators of $c/r_0 = 2.4$. From Fig.~\ref{fig1} we find that the corresponding effective interaction coefficient $K_z = -0.02$. In the bistable regime, when two stable uniform solutions exist in the chain, we observe the possibility of formation of switching waves, which represent a transition from one uniform distribution to another with a change in resonator number n. Such waves are initially excited by non-uniform distribution of the external field, and then they are supported by a uniform external excitation. The profiles of such waves are close to a step function, with typical domain wall structure shown in Fig.~\ref{fig3}(a).

\begin{figure}
\centering\includegraphics[width=13cm]{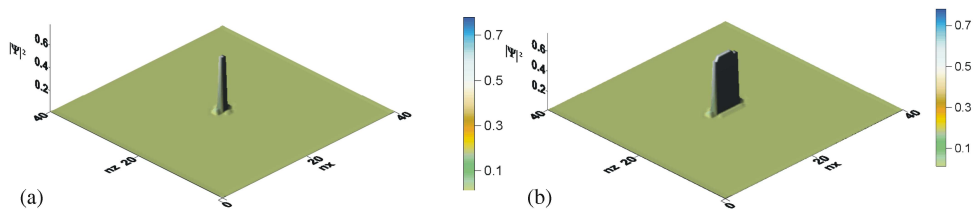}
\caption{Examples of a nonlinear localized mode in a two-dimensional lattice of SRRs. $\alpha=1$, $\Omega = -0.2$, $\gamma = 0.024$, $\Sigma = 0.058$.  (a) soliton with minimum possible size (b) – soliton with size 5x1. }
\label{fig4}
\end{figure}

Because of the discreteness of the system, the velocities and profiles of moving switching waves vary quasi-periodically with time (in contrast to a switching wave with a stationary profile in continuous media). For this reason, the velocity of the switching wave $v$ is obtained by averaging over sufficient time. The velocity sign is defined positive if the motion of the switching wave leads to the expansion of the
region occupied by the upper branch of bistability; otherwise, the velocity is negative. Figure~\ref{fig3}(b,c) shows the velocity of the switching waves as a function of the external excitation. We find that for the selected parameters there exist a narrow range of external excitations $\Sigma$ for which we observe bistable behavior, see Fig.~\ref{fig3}(c). Interestingly, one of the branches in bistable regime corresponds to the stationary domain wall, while another branch - to the one moving in negative direction.

\subsection{Dissipative solitons in two-dimensional lattices}

Now we study the two-dimensional arrays of the resonators, and we select the lattice orientation shown in Fig.~\ref{fig2}(c). We choose the normalized frequency of the excitation of $\Omega = -0.2$, and the spacings between resonators $a/r_0 =2.4$, $c/r_0 =1$. From the results presented in Fig.~\ref{fig1} we find the effective interaction coefficients for this regime as $K_x =  0.06$, $K_z =  -0.02$, and use these values in our further simulations of Eqs.~\ref{svea}. Our two-dimensional structure supports a family of nonlinear localized solutions -- discrete dissipative solitons. In bistable regime, when the interaction between resonators is sufficiently weak, such solutions of Eqs.~\ref{svea} can be found using perturbation theory, with the smallness parameter defined by the interaction coefficients. However for realistic interaction constants, shown in Fig.~\ref{fig1}, the existence and stability of dissipative solitons can be found only numerically. Fig.~\ref{fig4} shows two possible stable soliton solutions, consisting of just one strongly excited resonator (Fig.~\ref{fig4}(a)) or five excited resonators (Fig.~\ref{fig4}(b)). For such narrow solitons, as well as for narrow solitons in continuous media~\cite{Rosanov_book}, stability of the upper branches of the bistability curves is not generally required for the stability of the solitons, since the modulational instability is calculated for the homogeneously excited structure. However, for wider excitations, the modulational instability starts to manifest itself, with the larger structures showing instability and splitting into several smaller solitons. For example, if we initially excite the structure with wide area corresponding to the upper branch of the bistability curve (see Fig.~\ref{fig5}(a)), then the excitation becomes unstable in z-direction, and we observe formation of two stable narrow solitons, which are shown in Fig.~\ref{fig5}(b). Demonstrated decay of the wide soliton occurs within 100 dimensionless time units $\tau$.

\begin{figure}
\centering\includegraphics[width=13cm]{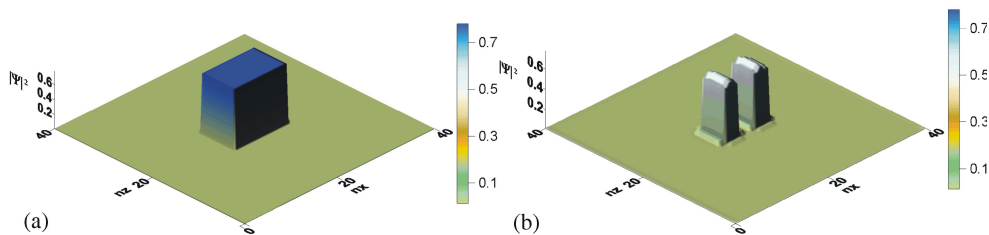}
\caption{(a) Initial and (b) final state of the two-dimensional magnetic domain in a SRR lattice in the case of anisotropic modulational instability. The domain decays in two identical modes of a smaller size. Parameters are: $\alpha =1$, $\Omega = -0.2$, $\gamma = 0.032$, $\Sigma = 0.063$.}
\label{fig5}
\end{figure}

\section{Conclusions}

We have analyzed nonlinear modes in one-dimensional arrays and two-dimensional lattices composed of weakly coupled split-ring resonators with both electric and magnetic coupling. We have demonstrated the existence of domain walls connecting two stable states of the metamaterial excited by an external electromagnetic field, and studied its dissipative dynamics characterized by the bistable velocity.
We have also studied two-dimensional nonlinear magnetic domains in nonlinear lattices of SRRs and demonstrated that larger domains may experience modulational instability.  Our results may be useful for the analysis of other types of nonlinear metamaterials
supporting subwavelength discrete localized modes, especially those operating in the optical regime, such as nanoscale periodic structures consisting of metal and nonlinear dielectric slabs where discrete solitons have been analyzed recently~\cite{zhang}.

\section*{Acknowledgements}

The authors acknowledge a financial support from a mega-grant of the Ministry of Education and Science of Russian Federation
and Australian Research Council, and they thanks G. Tsironis for useful discussions and comments.

\end{sloppy}
\end{document}